\documentclass[
 reprint,
 amsmath,amssymb,
 aps,
prb,
floatfix 
]{revtex4-2}
\usepackage{graphicx}
\usepackage{dcolumn}

\usepackage{bm}
\usepackage{hyperref}

\usepackage{braket}
\usepackage{csquotes}
\newcommand{\beq}{\begin{equation}}
\newcommand{\eneq}{\end{equation}}
\usepackage{xcolor}
\begin{document}

\title{Phase diagram of the disordered Kitaev chain with long-range pairing connected to external baths}
  
\author{Emmanuele G. Cinnirella$^{(1,2)}$, Andrea Nava$^{(3)}$, Gabriele Campagnano$^{(4)}$, and Domenico Giuliano$^{(1,2)}$}

\affiliation{
$^{(1)}$Dipartimento di Fisica, Universit\`a della Calabria Arcavacata di 
Rende I-87036, Cosenza, Italy \\
$^{(2)}$INFN, Gruppo collegato di Cosenza, 
Arcavacata di Rende I-87036, Cosenza, Italy \\
$^{(3)}$Institut f\"ur Theoretische Physik, Heinrich-Heine-Universit\"at, 40225 D\"usseldorf, Germany \\
$^{(4)}$CNR-SPIN,  c/o Complesso di Monte S. Angelo, via Cinthia - 80126 - Napoli, Italy}
 
\begin{abstract}
We  study  the  phase diagram of a disordered Kitaev chain with long-range pairing when 
connected to two metallic leads exchanging particles with external Lindblad baths. We (i) monitor the subgap modes at increasing disorder, (ii) compute  the  current flowing across the system at a finite voltage bias between the baths, and (iii) study the normal 
single particle lead correlations across the chain. Throughout our derivation, 
we evidence the interplay between disorder and  topology.  In particular, we evidence the reentrant behavior of the massive, topological 
 phase at limited values of the disorder strength, similar to what happens in the short-range pairing 
 Kitaev model. Our results suggest the possibility of a disorder-induced direct transition between the massive and 
 the short-range topological phase of the long-range pairing Kitaev model. 

\end{abstract}
\date{\today}
\maketitle

\section{Introduction}
\label{intro}
In the pursuit for systems with excitations suitable for quantum computation purposes, quantum topological systems have gained constantly increasing
importance.
The main feature of a quantum topological phase is the emergence of states localized at the boundary of the system which can only be removed by letting the system undergoing a topological phase transition\cite{Hasan2010,Qi2011}. 
Such boundary states in one dimensional (1D) systems  have been proposed to be useful in quantum computation protocols in several platforms, such as quantum Hall states at filling 5/2 \cite{Moore1991,Nayak2008}, helical states in electron liquids \cite{Oreg2010}, or in photonic metamaterials \cite{Mousavi2015} to mention a few. 

  A topological phase can be identified by a topological invariant  such as the Chern number ${\cal C}$---the integral of the Berry curvature over the Brillouin zone of the system with periodic boundary
 conditions \cite{Berry1984}. Typically, ${\cal C}$  is quantized and, due to the bulk-boundary correspondence\cite{Fidkowski2011,Chen2020,Zhong2024}, 
  its value, computed in the translationally invariant system with periodic boundary conditions, corresponds to the number of edge modes emerging at the boundaries of the system with open boundary conditions. In 
models with short-range interactions and/or electron hopping $|{\cal C}|=0$ or 1, respectively in the topologically trivial or nontrivial phases.  At variance, an hopping beyond nearest neighbors  can stabilize 
phases with $|{\cal C}|>1$, corresponding to multiple, 
independent, exponentially localized topological zero energy modes at the boundary of the system. Remarkably, this behavior, 
theoretically predicted for both normal
 \cite{Li2014,Li2019,Malakar2023,Gonzalez2019},
 and superconducting systems \cite{Li2021,Niu2012,DeGottardi2013,Lieu2018}, has recently been experimentally 
 observed in optical lattice simulations of topological systems \cite{Cardano2017,Derrico2020}.  

On the other hand, many experimental systems able to host topological phases of matter, such as Rydberg synthetic lattices 
\cite{Schauss2012,Kanungo2022,Meier2018}, optical waveguides \cite{Wang2022,Kang2023}, and trapped ions \cite{Grusdt2013}, 
exhibit power-law decaying interactions. Most relevant to our work,
power-law decaying interactions have been studied in the Kitaev chain with long-range tunneling and/or pairing hopping (LRK) 
 \cite{Vodola2016,Alecce2017,Patrick2017} and in the helical Shiba chain \cite{Pientka2013s,Pientka2014}, concerning 
  both static and dynamical properties in the presence of quenches 
\cite{Grass2014,Hauke2013,Dutta2017,Cevolani2018}, or periodic drivings \cite{Nandy2018,Bhattacharya2019}, as well as 
 the investigation of the entanglement entropy growth \cite{Koffel2012,Pezze2017}. Such interactions 
  are expected to lift the degeneracy of the zero energy modes, giving rise to 
 massive subgap topological states characterized by a finite energy, a wavefunction showing a power law spatial decay 
 inside the bulk and a noninteger  value of the topological invariant \cite{Olekhno2021,Kim2024,Gong2016,Medina2023}.
  
 In this work we address the phase diagram of the LRK in the presence of disorder. In 1D normal systems, any amount of disorder is known to fully localize the electronic 
wavefunctions, thus inducing a transition toward a disorder-induced insulating phase
\cite{Anderson1958, Abrahams1979,Evers2008}. At variance, 
in topological systems a limited amount of disorder may work to enforce the topological state, 
thus leading to a reentrant topological phase  
\cite{Zuo2022,Li2009,Liu2022,Nava2023ssh,Cinnirella2024}, till it  becomes strong enough 
to induce a phase transition to trivial phases. A reentrant topological phase is, by now, a well-established 
feature of 1D disordered, topological superconductors \cite{Pientka2013,Pientka2013,Nava2017,Zuo2022}. 
In the specific context of 
the LRK model, previous studies of the effects of disorder have concerned a closed system with a discommensurate
one-particle potential term \cite{Cai2017}  or a correlated commensurate one \cite{Fraxanet2021}, or the nonequilibrium 
dynamics following a parametric quench of the model Hamiltonian \cite{Mishra2020,Baghran2024}.  

  In general, disorder makes it difficult to even define ${\cal C}$, except in remarkable situations, where additional symmetries 
allow for consistently computing it \cite{Niu1984,Fedorova2020,Liu2022,Mondragon2014,Lin2021,Nava2023ssh,Fumina2024}. 
Thus, when investigating the combined effects of disorder and topology on the 
phase diagram of  the system, it is worth resorting to alternative means to set the boundaries of topological phases. 
In this direction, in Refs.\cite{Nava2023ssh,Cinnirella2024} we proposed to employ  
the even-odd differential occupancy of the lattice sites, combined with a systematic analysis of the transport properties of
the system, taken to a non-equilibrium steady state (NESS) by means of a coupling to an external dissipative (Lindblad) bath. Indeed, 
a direct coupling to the external  bath has already proven to 
be an effective tool to study  the phase diagram of topological  1D systems \cite{Nava2023ssh,Cinnirella2024}, 
as well as the onset of topological dynamical phase transitions
in two dimensional superconducting systems \cite{Nava2024S,Nava2024L}, and to detect the presence of non-trivial 
physical behaviors like the Mpemba effect \cite{Lu2017,Nava2024M}, the dynamical parity
 stabilization \cite{Ackermann2023,Zatsarynna2024} or the dissipative cooling \cite{Nava2022_2}. Along this line,   
  here we study the phase diagram of an open LRK with Anderson disorder  
\cite{Vodola2016,Alecce2017,Giuliano2018}, connected 
to  1D metallic leads and  allowed to exchange particles with a Lindblad bath at a given chemical potential
\cite{Lindblad1976,Pearle2012}. 

 Specifically, we implement three different means of identifying the various phases of the disordered 
 LRK: (i) we look  at the lowest and at the next-to-lowest energy mode 
 of the isolated chain, (ii) we monitor   the  current ${\cal I}$ across  the LRK, and (iii) the single-particle lead correlations across the LRK, $C_{L,R}$ in the NESS, when the baths are taken at a finite 
 voltage bias, as a function of both the potential bias between the 
 baths and the disorder strength. Our approach is  grounded on the different structure of the subgap modes in 
 the various phases of the LRK in the absence of disorder. Indeed, as we review in the following, the model shows
 a trivial phase, with no energy modes below the bulk gap, a topological phase, equivalent to the one of 
 the Kitaev model \cite{Kitaev2001}, with two degenerate, real fermionic modes pinned at zero energy (for a long 
 enough chain), and a ``topological massive phase'', with a finite-energy subgap mode, which can be regarded as
 being determined by an overlap between the real fermionic modes that keeps finite in the infinite chain limit \cite{Vodola2016,Alecce2017,Giuliano2018}.  Doing so, we  evidence the robustness of 
the nontrivial phase with  a finite energy subgap mode (a typical feature of the LRK 
 \cite{Viyuela2016,Lepori2016,Bhattacharya2018} which is related to the possibility, for the
  LRK, to work as an efficient “particle entangler”, 
 by means of a mechanism dual to the crossed Andreev reflection across the Kitaev chain \cite{Giuliano2018}) 
 against Anderson disorder.
 
  Throughout our derivation, we first show how to identify the various phases of the LRK in the absence of disorder
  by looking at the low-lying energy modes of the system, at ${\cal I}$ and at $C_{L,R}$.
  Then, we compute them in the presence of disorder  and monitor up to which value of the disorder 
 strength a  feature characterizing a given phase survives (which is a signal that the corresponding phase is still surviving the disorder),
 setting the corresponding phase boundary at the point where the feature itself disappears. In this way, we are able to  
  map out the  phase diagram  of our system. Eventually, we 
 verify that the results independently obtained by looking at the three quantities are all consistent with each other.

Our paper is organized as follows:

\begin{itemize}

\item In Sec. \ref{modham}, we present the model Hamiltonian for the LRK  in the  
 absence of disorder and briefly review its phase diagram. Then, after adding
 Anderson disorder, we look at the two lowest-lying energy eigenvalues to argue how, and to what extent, the disorder modifies
  the phase diagram, with respect  to the clean limit.

\item In Sec. \ref{lindblad}, we introduce the Lindblad master equation approach to   the LRK  coupled to two 
external leads exchanging particles with Lindblad 
baths at different potentials and recover the formulas for computing ${\cal I}$ and $C_{L,R}$ in the 
NESS. 

\item In Sec. \ref{int_cur},  we study the behavior of the current and of the modulus of the 
  endpoint correlations between the metallic leads in the NESS and relate it to the   the various phases of the disordered LRK.

\item  In Sec. \ref{concl}, we summarize our main conclusions and present possible further developments of our
work.

\item In Appendix, we provide mathematical details of the formalism 
of Lindblad Master Equation that we employ throughout our derivation.  

\end{itemize}

\section{Model Hamiltonian}
\label{modham}

In the following,    we review  the model Hamiltonian and the phase diagram of the isolated  LRK, first in the  
clean limit, and then at nonzero disorder strength.
 
\subsection{The long-range Kitaev chain in the absence of disorder}
\label{LRK_nod}

\begin{figure}[t]
  \centering
  \includegraphics[width=\linewidth]{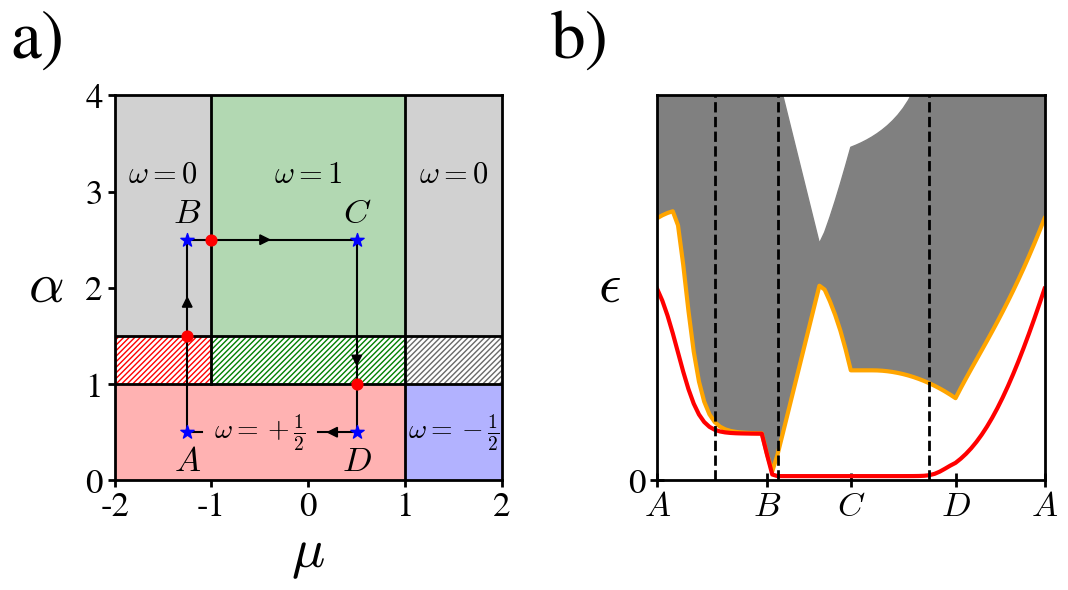}
  \caption{{(a)}:  Phase diagram of the LRK described by $H$ in Eq. (\ref{eq:LRK}) with $2w=\Delta=1$ and $L=500$, 
  drawn in in the $\mu$-$\alpha$ plane. The various colors identify phases with   different values of $\omega$  (see the 
  main text for details). 
  Hatched areas correspond to crossover regions, where the winding number interpolates between the adjacent regions. 
  The red dots mark the transition points between phases with distinct topological properties. \\
{(b)}: Spectrum of $H$ along the curve ${A\rightarrow B\rightarrow C\rightarrow D\rightarrow A}$ highlighted in
{(a)},   with $\mu_A=\mu_B=-1.25$, $\mu_C=\mu_D=0.5$, $\alpha_A=\alpha_D=0.5$ and $\alpha_B=\alpha_C=2.5$.
   The solid red line corresponds to the lowest energy level, the grey filled areas represent window of allowed bulk energy levels, 
 the solid orange line corresponds to the value of the gap $E_\Delta$,  the dashed black 
   lines highlight the crossing from one
   phase to another, signaled by the red dots in {(a)}.
      }
  \label{fig:ps_clean}
\end{figure}

The   LRK describes spinless fermions over a one-dimensional (1D) lattice, with a nearest neighbor normal
hopping strength $w$ and a \textit{p}-wave, superconducting pairing term decaying, in real space, as a power of the 
distance between two sites. Over an $L$-site chain, the LRK Hamiltonian $H$ is given by    \cite{Vodola2014,Lepori2016,Vodola2016,Giuliano2018} 

\begin{equation}
\begin{aligned}
    H
    =& -w\sum_{j=1}^{L-1}\left(\chi^\dagger_j\chi_{j+1}+\chi^\dagger_{j+1}\chi_j\right)-\mu\sum_{j=1}^{L}\chi^\dagger_j\chi_{j} \\
    &+\frac{\Delta}{2}\sum_{j=1}^{L-1}\sum_{r=1}^{L-j}\delta_r^{-\alpha} \left(\chi^\dagger_j\chi^\dagger_{j+r}+\chi_{j+r}\chi_j\right)
    \:\: . 
\end{aligned}
\label{eq:LRK}
\end{equation}
\noindent
In Eq. (\ref{eq:LRK}) $\chi_j$ ($\chi_j^\dagger$) are  the annihilation (creation) operator for a spinless fermion at site 
$j$ of the chain. They satisfy the standard anticommutation relations 
$\{\chi_j , \chi_{j'}^\dagger \} = \delta_{j,j'}$, $\{\chi_j , \chi_{j'} \} =
\{\chi_j^\dagger , \chi_{j'}^\dagger \} = 0$. $w$ is the single-fermion 
hopping strength between nearest-neighboring sites, $\mu$ is the chemical potential, $\Delta$ is the pairing strength
and $\delta_r =|r|$ if $j+r \leq L$, otherwise $\delta_r = 0$. 
The exponent $\alpha$ determines the ``effective range'' of the interaction. $\alpha \to \infty$ corresponds to the 
``standard'' Kitaev model, with \textit{p}-wave pairing between nearest-neighboring sites over the lattice \cite{Kitaev2001}. 
In the following, for the sake of simplicity (and without loss of generality), 
we set   $\Delta=2w=1$, to reduce the number of parameters.
Switching to momentum basis and employing Nambu spinor representation,  we rewrite  $H$ in Eq. (\ref{eq:LRK}) as 

\beq
    H=\sum_k 
    [\chi_k^\dagger , \chi_{-k} ] [\vec{h}_{\alpha,\mu} (k ) \cdot \vec{\sigma} ] \left[ \begin{array}{c} \chi_k \\ \chi_{-k}^\dagger 
    \end{array} \right] 
    \:\: , 
    \label{matrixform} 
 \eneq
 \noindent
 with $ \chi_k  =\frac{1}{\sqrt{L}}\sum_{j=1}^L \: e^{ikj}\chi_j$, $\vec{\sigma}$ being the Pauli matrices, 
and \cite{Vodola2014,Lepori2016}
 
 \beq
 \vec{h}_{\alpha,\mu} (k) \cdot \vec{\sigma}  = \left[ \begin{array}{cc} -\cos(k)-\mu & - i f_\alpha (k+\pi) \\ 
 -i f_\alpha (k+\pi) & \cos (k) + \mu \end{array} \right] 
  \:\: ,
 \label{vech}
 \eneq
 \noindent
  with  $f_\alpha (k) = \sum_{r=1}^{L-1} e^{ikr} /\delta_r^\alpha$. 
  The model Hamiltonian in Eq. (\ref{eq:LRK}) belongs   to the BDI class of the topological insulators and
  superconductors \cite{Atland1997}.  
  Its (bulk) quasiparticle spectrum is determined by the dispersion relation $E_k = \sqrt{(\cos (k) + \mu)^2 + |f_\alpha(k)|^2}$.
 Out of the critical lines $E_k$ takes a finite gap $E_\Delta$, which corresponds to the orange, solid line in 
 Fig. \ref{fig:ps_clean}{(b)}. 
 
The topological phases of an Hamiltonian in the form of Eq. (\ref{matrixform}) is typically signaled
by a nonzero value of the winding  number $\omega$ (which, in 1D systems coincides with ${\cal C}$), given by  \cite{Zak1989}

\beq
\omega = \frac{i}{\pi} \int_{\rm BZ} \: d k \: \langle u_k | \partial_k | u_k\rangle 
\:\: , 
\label{wn.1}
\eneq
\noindent
with $|u_k\rangle$ being the negative-energy eigenstate of $h_{\alpha,\mu} (k)$, and 
the integral performed over the full Brillouin zone \cite{Tewari2012}.  In the case of 
a short-range Hamiltonian,  $\omega$ measures how many times  $\vec{h}_\alpha (k)$ revolves around the origin
as $k$ spans the whole Brillouin zone and, therefore, it can only take integer values. In our specific cases, 
  we point out that, for $\alpha <1$,  the function $f_\alpha (k)$ in Eq. (\ref{vech})  
  has a singularity for $k=0$, which gives rise to a  third, ``intermediate'' phase, 
  in addition to the trivial and to the topological one of the Kitaev model \cite{Viyuela2016,Lepori_2017,Bhattacharya2018},
  as well as  of crossover regions, corresponding to the hatched areas of  
 Fig. \ref{fig:ps_clean}{(a)}. In the crossover regions $\omega$ is ill defined, thus losing its 
direct connection with the topological properties of the phase   (see
 Ref.\cite{Lepori_2017} for a detailed discussion of this point). 
 
 In Fig. \ref{fig:ps_clean}{(a)} we draw the phase diagram of the LRK  in the $\mu$-$\alpha$ plane.   
 In particular,  we identify:

\begin{itemize}

    \item {\bf For $\alpha>3/2$ and $|\mu| < 1$}, a topologically nontrivial phase [green region of Fig. \ref{fig:ps_clean}{(a)}], 
    corresponding to the topological phase of the short-range Kitaev model and, as such, characterized by 
    the winding number $\omega =1$. In this phase the system exhibits a    gapped spectrum and a pair of real-fermionic
    zero modes localized at the endpoints of the chain, whose overlap is exponentially suppressed with the length 
    of the chain itself.  
    
    \item {\bf For $\alpha>3/2$ and $|\mu|>1$}, a topologically trivial phase [grey regions of Fig. \ref{fig:ps_clean}{(a)}], 
with $\omega = 0$, corresponding  to the 
    trivial phase of the short-range Kitaev model \cite{Vodola2014}). In this 
    phase  the system is gapped and  no edge state is present.

    \item {\bf For $\alpha<1$ and $\mu>1$}  [blue region of Fig. \ref{fig:ps_clean}{(a)}],  
    there are no subgap topological states, which makes this region similar to the trivial one, for what
    concerns its topological properties  \cite{Bhattacharya2018}.

    \item {\bf For $\alpha<1$ and $\mu<1$} [red region of Fig. \ref{fig:ps_clean}{(a)}], the system is   gapped 
    with localized  real fermionic modes at the edges. Differently from the topological phase, 
    now there is a finite overlap between the isolated edge modes, decaying as a power-law of the length of the chain. 
    Accordingly, they hybridize and give rise to massive excitations and, at the 
    same time, one gets a noninteger winding number $\omega=1/2$, as 
  a consequence of the singularity in both the dispersion relation and in the group velocity \cite{Bhattacharya2018}.
 In the following we refer to this phase as to the ``massive topological phase.'' 
     
        \item {\bf For $1<\alpha<3/2$ and $|\mu|<1$}  [hatched green area of Fig. \ref{fig:ps_clean}{(a)}],  
    there is still a subgap, zero energy modes. However, the winding number is ill defined, due to the 
    divergence of the group velocity at $k=0$  \cite{Bhattacharya2018}.

        \item {\bf For $1<\alpha<3/2$ and $ \mu <-1$}  [hatched reddish  area of Fig. \ref{fig:ps_clean}{(a)}],  
   the behavior of the system is the same as in the above region, except that the subgap mode is now massive.

\item Finally, for  {\bf For $1<\alpha<3/2$ and $\mu>1$} [shaded grey area of Fig. \ref{fig:ps_clean}{(a)}],
 the system undergoes a crossover between
two phases, both topologically trivial. 

\end{itemize}
\noindent

\subsection{The disordered long-range Kitaev chain} 
\label{sec:disorder} 

\begin{figure}[t]
  \centering
  \includegraphics[width=\linewidth]{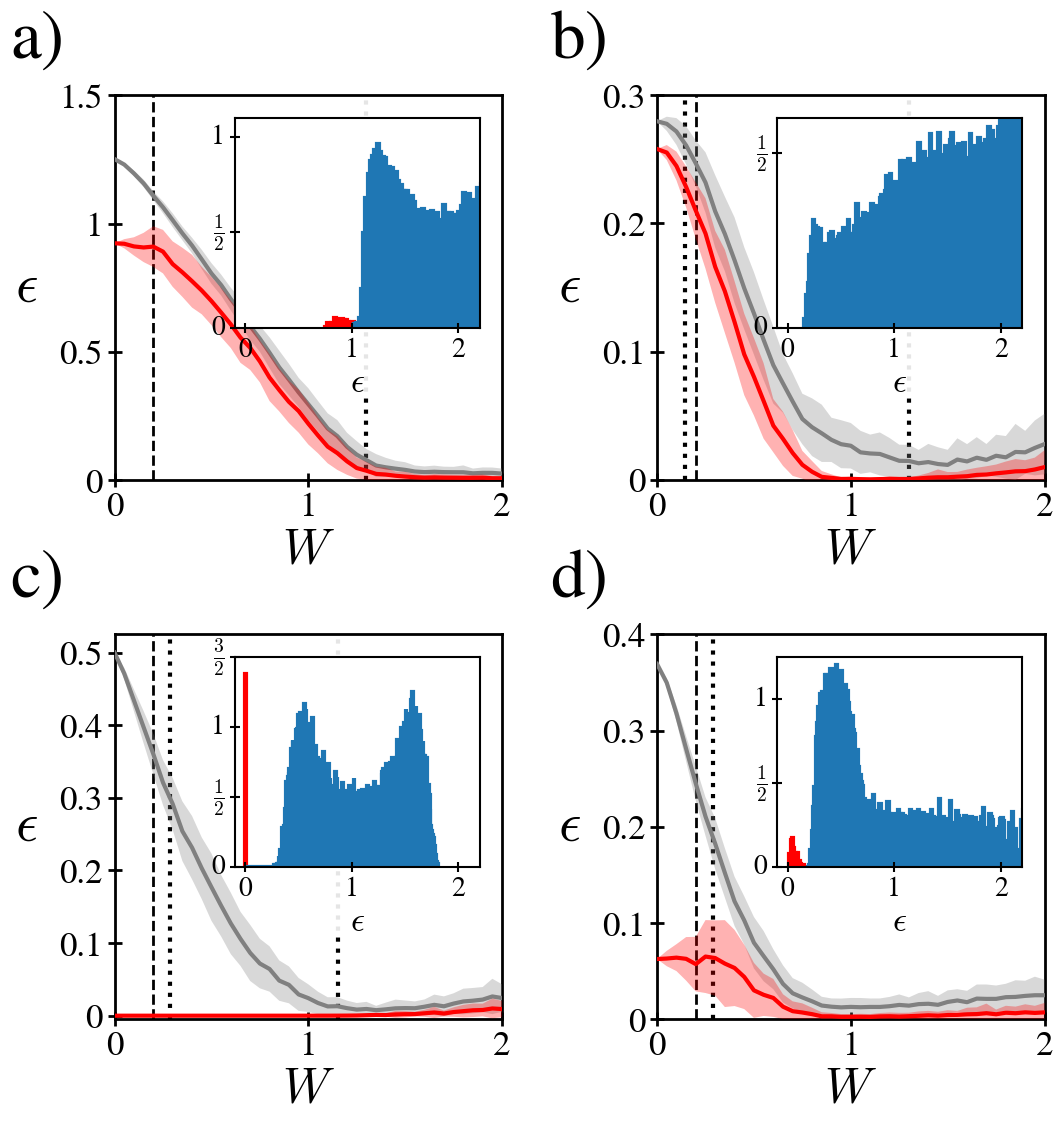}
  \caption{Lowest energy   (red line) and next to the lowest energy levels (grey line) for the disordered LRK as    function of the disorder
   strength $W$ corresponding to the point A ({a)}), B ({b)}), C ({c)}) and D ({d)}) of Fig. \ref{fig:ps_clean}. The
plots have been drawn for 
$L=500$ by averaging over $100$ different disorder configurations for each fixed $W$. The shaded  regions 
measure the corresponding standard deviation at a given $W$. Inset plots: numerical energy density of the full 
spectrum in the range $\epsilon\in[0,2]$, computed using the same disorder configurations of the main plots, at the value of
$W$  corresponding to the black dashed lines in the main plots, with the peaks corresponding to 
the subgap modes highlighted in red.  }
  \label{fig:spectrum_disorder}
\end{figure}

To introduce impurity disorder, we  add to the chemical potential $\mu$ 
a random, site-dependent additional term $\epsilon_i$, that is, we set $\mu_i = \mu \longrightarrow \mu_i = \mu + \epsilon_i$,
with  $\{\epsilon_i\}$ being random numbers, distributed according to the probability distribution  

\beq
P[\epsilon]=
    \begin{cases}
        \frac{1}{2\sqrt{3}W} & \text{ if } |\epsilon|\le\sqrt{3}W \\
        0 & \text{ otherwise}
    \end{cases} \;\; , 
\label{eq:rand_distr}
\end{equation}
\noindent
and the ``disorder strength'' $W (>0)$  defined so that $W^2$ corresponds to the variance of $P[\epsilon]$. 
Disorder, as we introduce it, breaks the lattice translational invariance. Thus, being able to provide 
a generalization of the winding number to the disordered case is, in general, no longer simple 
and typically depends on the specific 
system one is considering   \cite{Phillips1991,Zuo2022,Nava2023ssh}. 
For this reason, to identify the various phases of the system,   in the following  
we  look at how disorder itself affects the  subgap modes in the LRK spectrum, motivated by 
  our above observation that  the various
phases can be identified by monitoring those  modes and how 
their energies behave as a function of the system parameters \cite{Lepori2016,Giuliano2018}.  We first compute the energies of the first two low-lying levels of 
the system at a given value of the system parameters and for a fixed configuration of disorder, then we average over the disorder
realizations, generated by using the probability distribution in Eq. (\ref{eq:rand_distr}). 
To double check whether we recover  a faithful description of   the disordered system, we also compute the standard deviation
of the energies from the average. Doing so as a function of the disorder strength $W$, we recover the plots shown in 
 Fig.  \ref{fig:spectrum_disorder}.

\noindent
There, we plot,  with two shaded red and grey curves,
 the disorder averaged value of the  lowest and of the next to lowest single particle energies as a function of 
$W$, computed at the   points A,B,C, and D in parameter space  highlighted in the plots in Fig. \ref{fig:ps_clean}. 
Moreover, the shaded width of both curves measures  the standard deviation of the energies 
around their average value. 
As long as the shaded curves do not   overlap with each other,
 we assume that the integrity of the levels is preserved. Apparently, up to   $W \approx 1$, 
 we see that the various phases are not 
 sensibly affected by disorder.  On further increasing $W$,  the system enters   a disorder dominated phases, whose 
specific features depend on  the corresponding phase in the clean limit. Specifically,  if, in the clean limit, 
the system is in the topologically trivial phase [Fig. \ref{fig:spectrum_disorder}{(b)}],  increasing the disorder pushes the lowest, single-particle 
energy toward zero, while the standard deviation keeps rather low and constant, compared to the energies. Then, 
a ``turning point'' is reached around $W \approx 1$,   after which the standard deviation starts to increase.
This is a signature of a
reentrant topological phase for intermediate disorder strength, similar
to the one predicted in the Kitaev model with short-range pairing \cite{Nava2017}.  
In  Fig. \ref{fig:spectrum_disorder}{(c)} we show the effects of the disorder when, in the clean limit, the system is in 
the nontrivial topological phase, with $\omega =1$. In this case, there is a  subgap  state at exactly
zero energy, encoded in the two unpaired Majorana modes at the endpoints of the chain \cite{Kitaev2001}. Accordingly, 
we see that, at low enough values of $W$,  the lowest-energy eigenvalue is pinned
at zero energy, with a negligible spread of the values at a given disorder strength (witnessed by the fact that, for low $W$, 
the red plot is just a sharp line). The natural conclusion is a disorder-enforced stability of the zero-energy fermionic mode, 
consistently with the plot in the inset of Fig. \ref{fig:spectrum_disorder}{(c)}, where we report the density of states 
$\rho (\epsilon)$ at a representative value of $W$. The sharp, subgap peak at zero energy is the apparent counterpart of the 
(weak) disorder robust, subgap level at $\epsilon = 0$. At variance, at large enough values of  $W$, the grey region overlaps with 
the red one: this corresponds to the disappearance of the zero-energy sharp resonance in the density of states and, ultimately, 
to the disappearance of the topological phase.

 Figures \ref{fig:spectrum_disorder}{(a)} and \ref{fig:spectrum_disorder}{(d)} correspond to the massive topological phase. Indeed, close to 
 the clean limit (low values of  $W$), the red and the grey lines are well separated from each other, by an amount that, on the average, is of the order of the 
 energy difference between the subgap state and the first bulk state over the gap. As $W$ increases, both energies monotonically
 decrease, on the average,  toward zero. At a critical value of $W_c$, which we estimate below, the two curves 
 fuse with each other, thus signaling the disappearance of the massive topological phase, exactly like in the plot of 
  Fig. \ref{fig:spectrum_disorder}{(c)}. 
  For $W>W_c$, both the average and the standard deviation of the energies further increase with $W$, similarly to
 what happens in  Fig. \ref{fig:spectrum_disorder}{(c)}. By analogy, we conclude that, in all three cases, such a behavior is a signal
 that the system has undergone a disorder-induced transition toward a topologically  trivial phase.

  To  estimate $W_c$, we note that, at any given $\alpha$ and $\mu$, on a 
  single realization of the disorder, the chemical potential at each site lies 
  between  $\mu_<=\mu - \sqrt{3} W$ and $\mu_> = \mu + \sqrt{3} W$. It follows that, while the average 
  chemical potential of the disordered system, 
  in the thermodynamic limit, is still expected to be equal to  $\mu$, increasing $W$ increases also the 
  probability to randomly generate finite 
  segments of the chain ("islands") characterized by an average chemical potential $\mu_{\ell}$ different 
  from $\mu$. Such islands are responsible for the proliferation of 
  many low energy subgap states. Depending on $\mu$ and on the disorder strength $W$, the 
  islands can be characterized by a value of $\mu_{\ell}$ 
  that belongs to a different phase compared to the clean system and can, in principle, drag the system across a topological phase 
  transition. If we consider,
   for example, a clean system in the topological nontrivial phase ($\alpha>3/2$, $|\mu|<1$), such as the one corresponding to point C of 
   Fig. \ref{fig:ps_clean}{(a)}, the formation of islands belonging to the topological trivial one ($|\mu_{\ell}|>1$) 
   will generate low energy states that can hybridize with the zero energy  modes destroying them, thus driving the disordered 
   system into the topological trivial phase. 
   On the contrary, topological nontrivial island ($|\mu_{\ell}|<1$), in a topological trivial background ($\alpha>1$, $|\mu|>1$), 
   can give rise to a reentrant topological phase.

The probability of formation of such topological nontrivial (trivial) islands in a trivial (nontrivial) background depends 
on the number of sites having a chemical 
potential lying beyond the transition lines of the clean system. The same argument  applies also to the topological massive case. 
We can define $W_c$ as 
the minimum amount of disorder at which the chemical potential over some sites starts to lie outside the boundary of the original phase. Eventually, we define
 $\tilde{W}_c$ as the critical value at which the probability is maximum. Apparently, $W_c$ and $\tilde{W}_c$ depend on the two phases between 
 which the disorder induces a transition. 
 
 To conclude this Section, we point out that, while the bulk gap provides a mean to identify the phases 
 of the LRK in the presence of disorder, its  magnitude   is not an indicator of a disorder induced phase transition, 
 as the disorder may further localize the zero-energy mode eigenfunctions, resulting in the closing of the energy 
 gap without a phase transition in the bulk of the system\cite{Niklas2016,Nava2017,Li2021}. Therefore,   to exactly locate the disorder-induced
 transition line, one has to resort to additional methods \cite{Niklas2016}, such as looking at the current ${\cal I}$ and at
 the single-particle lead correlations $C_{L,R}$, which we do in the following of our paper.

\section{Lindblad Master Equation approach to the open LRK model}
\label{lindblad}

\begin{figure}[t]
  \centering
  \includegraphics[width=\linewidth]{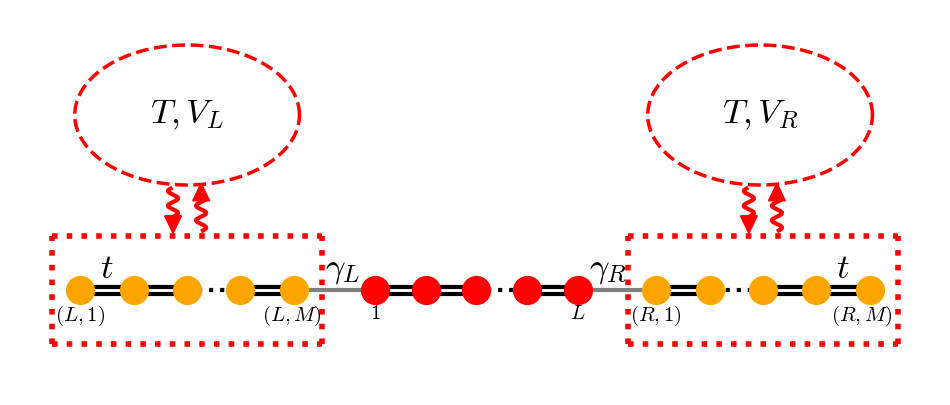}
  \caption{Sketch of the LRK (red-dot chain)  with the metallic leads (orange-dot chains) undergoing particle 
  exchange processes  (represented by the wavy red lines) with  
the thermal baths (reservoirs). The reservoirs (represented by the red dashed ovals)
are kept at temperature $T$ and at voltage bias (with respect to the LRK) 
respectively given by $V_L$ (left reservoir) and by $V_R$ (right reservoir).  }
  \label{sktch}
\end{figure}

In  Fig. \ref{sktch}, we sketch the LRK with the metallic leads, connected to the external
reservoirs.  We assume a thermal  distribution of the fermionic modes
in each reservoir and a voltage bias between the two of them, with the left (right) reservoir
kept at potential $V_L$ ($V_R$), with respect to the LRK. To describe the time evolution of 
the whole system toward the NESS, we     employ the method of 
Refs. \cite{Guimaraes2016,Nava2021,Nava2023ssh,Cinnirella2024}.  In particular, assuming 
 that the bath relaxation time is much shorter than 
the timescales associated to the system (LRK plus leads), we resort to  Markov approximation
 \cite{Breuer2007,Weiss2021}. Systematically integrating over the bath degrees of freedom, we recover
 the Lindblad master equation (LME) approach to the open quantum system \cite{Lindblad1976,Pearle2012,Guimaraes2016,Nava2021,Nava2023ssh,Cinnirella2024}. In particular,
 to make the system evolve toward a NESS in which a finite bias is applied to the LRK via the metallic leads,
 we follow the approach of     Ref.\cite{Guimaraes2016} and construct the LME so to let the reservoirs   directly 
 exchange particles with the leads.

\noindent 
In the Markovian limit, the time evolution equation for the density matrix $\rho (t)$ of a system described by the Hamiltonian 
${\cal H}$, 
connected to the external leads 
coupled to the reservoirs,  is determined by the Liouville equation, given by   \cite{Lindblad1976,Pearle2012}:

\beq
   \frac{ d\rho(t)}{dt} =-i\left[ {\cal H},\rho(t)\right]+\sum_k\left(L_k\rho(t)L^\dagger_k-\frac{1}{2}\left\{ L^\dagger_k L_k,\rho(t)\right\}\right)
\label{eq:Lindblad}
\:\: . 
\eneq
\noindent
In our case,  ${\cal H}$ in Eq. (\ref{eq:Lindblad})  coincides with  the 
Hamiltonian governing the unitary evolution of the isolated system (LRK plus leads). 
The {\it jump operators} $\{L_k\}_k$  
encode the interactions between the reservoirs  and the system. 
To provide their explicit expressions, we model both leads 
(denoted, in the following, respectively with the labels 
$_L$ and $_R$) as noninteracting fermionic chains at zero chemical potential, described  by  $M$-site lattice Hamiltonians 
for spinless fermions, $H_{(L,R)}$ given by 

\begin{equation}
    H_{(L,R)}=-t\sum_{i=1}^{M-1}\: \{ d^\dagger_{(L,R),i}d_{(L,R),i+1}+d^\dagger_{(L,R),i+1}d_{(L,R),i}\}
    \:\: ,
\label{eq:H_met}
\end{equation}
\noindent
with $t$ being the nearest-neighbor hopping amplitude. The corresponding eigenenergies and eigenmodes, $\epsilon_{n}$, 
$\eta_{(L,R),n}$, are given by 

\begin{eqnarray}
&& \epsilon_n = - 2t \cos (k_n) \;\; (k_n=\pi n/(M+1) ) \nonumber \\
&& \eta_{(L,R),n} =  \sqrt{\frac{2}{M+1}} \: \sum_{j=1}^M \: \sin (k_n j) \: d_{(L,R),j} 
\:\: . 
\label{specific}
\end{eqnarray}
\noindent
The leads are connected to the LRK via the hopping term 
 $H_H = H_{H,L}+H_{H,R}$, with

\begin{eqnarray}
H_{H,L} &=& - \gamma_L \{d_{L,M}^\dagger \chi_1 + \chi_1^\dagger d_{L,M} \} \nonumber \\
H_{H,R} &=& - \gamma_R \{ d_{R,1}^\dagger \chi_L + \chi_L^\dagger d_{R,1} \}
\:\: , 
\label{hopping}
\end{eqnarray}
\noindent
with $\gamma_{(L,R)}$ being the hopping amplitudes. Finally, to stabilize a NESS 
with lead modes distributed according to  Fermi distribution, we assume that 
the external baths interact with the metallic leads by injecting and/or extracting 
eigemodes, at a corresponding rate proportional to the Fermi distribution function 
determined by the temperature $T$ (which we assume to be the same in all the 
reservoirs) and by the voltage bias \cite{Nava2021,Nava2024S,Nava2024L}. 
Accordingly, we define the following set of Lindblad jump operators, respectively
corresponding to the injection and to the extraction of lead excitations \cite{Guimaraes2016}:

\begin{eqnarray} 
    L^{\rm in}_{(L,R),k} &=& \sqrt{2\gamma_{(L,R)}  f_{(L,R)} (\epsilon_k) }\eta^\dagger_{(L,R),k} \nonumber \\
    L^{\rm out}_{(L,R),k} &=& \sqrt{2\gamma_{(L,R)}  (1-f_{(L,R)} ( \epsilon_k) ) }\eta_{(L,R),k} 
    \:\: , 
    \label{jump.1}
    \end{eqnarray}
    \noindent
    with the Fermi distribution function

\beq    
f_{(L,R)} (\epsilon_k)  =\frac{1}{1+\text{exp}\left(\frac{\epsilon_k-V_{(L,R)} }{T}\right)}
\:\: , 
\label{eq:jump_operators}
\eneq
\noindent
and $V_{(L,R)}$ being the bias of the reservoir connected to the $(L,R)$ lead. Specifically, 
in the following we assume  $V_L=-V_R = V (>0)$, which  
implies that the current flows from the left to the right, and $T=0$. 

To compute ${\cal I}$ and $C_{L,R}$ we need the (time-dependent)   correlation matrix elements,
$\theta_{i,j}$ and $\theta^A_{i,j}$, defined as

\begin{eqnarray} 
    \theta_{i,j} (t) &=& \text{Tr}\left( c^\dagger_i c_j \rho (t) \right) \:\:,\\
    \theta^A_{i,j} (t) &=& \text{Tr}\left( c^\dagger_i c^\dagger_j \rho (t) \right)
\:\: , 
\label{eq:corr_def}
\end{eqnarray}
\noindent
with the operators $c_i$ given by
\begin{equation}
c_i=
\biggl\{ \begin{array}{l} 
    d_{L,i} \text{ for } i\in[1,M] \\
    \chi_{i-M} \text{ for } i\in[M+1,M+L] \\
    d_{R,i-M-L} \text{ for } i\in[M+L+1,2M+L]
\end{array}
\;\; . 
\label{eq:conv_NSN}
\end{equation}
\noindent
The diagonal elements $\theta_{i,i} (t)$ correspond to the
 average occupation of each site, while the off-diagonal elements are related to the normal correlation between different sites. 
Conversely, $\theta^A (t)$ encodes the anomalous correlations between sites. 
 If (as it is in our case) the overall Hamiltonian is quadratic in 
 the fermionic operators and the Lindblad jump operators are linear in the fermionic modes, 
it is possible to write a closed set of linear differential equations for the time evolution of the $\theta_{i,j} (t)$ and of the $\theta_{i,j}^A (t)$. 
To do so, we rely on the formalism presented in Appendix  \ref{sec:appendix_current}, whose starting point consists in rewriting the 
system Hamiltonian in terms of the lattice fermion operators $c_j$ in Eq. (\ref{eq:conv_NSN}) as

\beq
 {\cal H} =\sum_{i,j} \left( A_{i,j} c^\dagger_i c_j + B_{i,j} c^\dagger_i c^\dagger_j +  B^*_{j,i} c_i c_j \right)
\;\; . 
\label{appe.1.1.t}
\eneq
\noindent
$A$ and $B$ in  Eq. (\ref{appe.1.1.t}) are 
$(2M+L)\times(2M+L)$ square matrices. Writing ${\cal H}$ as in Eq. (\ref{appe.1.1.t}) allows us 
to write in a compact form the time evolution equations for the    generalized covariance matrix $\Theta (t)$,  given by 

\beq
\Theta (t) = \left[ \begin{array}{cc} \theta (t) & \theta^A (t ) \\ (\theta^A (t))^\dagger & - \theta^T (t) \end{array}
\right]
\;\; . 
\label{def.1}
\eneq
\noindent
Within LME approach, the time evolution 
equation for $\Theta (t)$ is recovered by solving the differential equation

\beq
\frac{d \Theta (t)}{dt} = i [\Omega , \Theta (t)] -
\frac{1}{2} \{ \Gamma + \Pi , \Theta (t) \} + \Gamma 
\:\: ,
\label{def.5}
\eneq
\noindent
with the matrices $\Omega , \Gamma , \Pi$ determined by ${\cal H}$ and by the Lindblad jump operators,  and
 respectively defined as   

\beq
\Omega = \left[ \begin{array}{cc} A^T & -2B^\dagger \\ -2B & - A \end{array} \right]
\:\: , 
\label{def.2}
\eneq
\noindent

\beq
\Gamma = \left[ \begin{array}{cc} {\cal G} & - 2 i B^\dagger \\ 2iB  & - {\cal G} \end{array} \right]
\;\; , 
\label{def.3}
\eneq
\noindent

\beq
\Pi = \left[\begin{array}{cc} {\cal R} & 2iB^\dagger \\ -2iB & 2 {\cal G} + {\cal R}
\end{array}
\right]
\:\: . 
\label{def.4}
\eneq
\noindent
 We refer to   Eq. (\ref{appe.1.9}) of 
 Appendix  \ref{sec:appendix_current} for the rigorous definition of 
 these matrices and for a discussion of their properties.  
 
Equation (\ref{def.5}) readily allows us to derive  $\Theta$ at the NESS, $\Theta_{\rm NESS}$.
Indeed,  from the condition  $\frac{d \Theta_{\rm NESS}  (t)}{d t}  =0$,  we find 

\beq
{\cal A} \Theta_{\rm NESS} + \Theta_{\rm NESS} {\cal A}^\dagger = - \Gamma
\;\; , 
\label{def.6}
\eneq
\noindent
with 

\beq
{\cal A} = -\frac{1}{2} \{\Gamma + \Pi \} + i \Omega 
\:\: . 
\label{def.7}
\eneq
\noindent
Equation (\ref{def.5}), as well as the condition for recovering $\Theta_{\rm NESS}$, Eq. (\ref{def.6}), are 
the key ingredients for our following calculation of the  charge current  and of the 
correlations.
 
\section{Charge current and endpoint correlations between the leads in the non equilibrium steady state}
\label{int_cur}

Coupling the system to the reservoirs as we discuss before stabilizes a NESS in which the leads 
are biased at a voltage $\pm V$ with respect to the LRK. This induces a finite current ${\cal I}$ through the 
system. In the clean limit, the behavior of ${\cal I}$ as a function of $V$ depends on whether 
the LRK develops subgap modes and whether they are massless, or massive. In particular, as we argue in 
the following within the framework
of a simplified model, in which a single, massless or massive mode is coupled to the two leads, 
${\cal I}$ as a function of $V$ shows an activated behavior at the energy of the mode itself. In addition, 
no activation is apparently expected if the system lies within   the topologically trivial phase,  
consistently with the absence of a subgap mode in this phase. 
A similar behavior is shown by the single particle lead correlations across the LRK. Importantly, 
the activated behavior survives in the presence of disorder, though with minor modifications. 
This, as we show below, provides us with an effective way to probe the 
 phase diagram of the LRK in the presence of disorder.

 \subsection{Single-level model as a reference calculation}
 \label{slm}
 
\begin{figure}[t]
  \centering
  \includegraphics[width=0.65\linewidth]{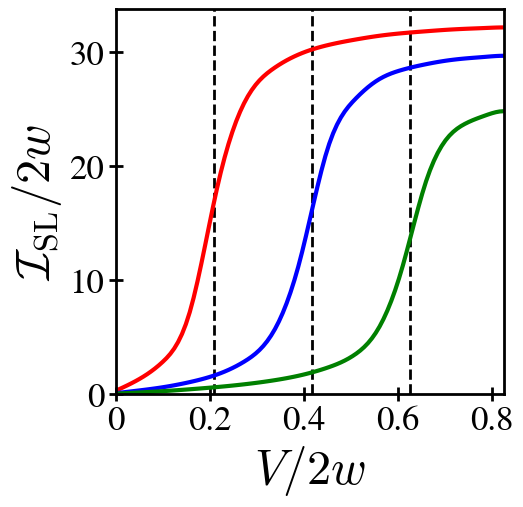}
  \caption{Current ${\cal I}_{\rm SL}/2w$  in the NESS, computed according to Eq. (\ref{appe.2.3}), 
  for the system described by the Hamiltonian in Eq. (\ref{appe.2.1}), with the following values of the various parameters:
 $2w=1$, $\tau=0.04$,  $\epsilon_0=0.21,0.42,0.63$ (red, blue and green curve, respectively). The dashed, vertical lines
 mark the position of the values of $\epsilon_0$ we used in the calculations.  [Note in this, and in following plots, 
 the dimensionless quantities reported on the ordinate axis have been multiplied by $10^4$, to ease the readability of 
 the figure.] }
  \label{f.appe.2.1}
\end{figure}

\begin{figure}[t]
  \centering
  \includegraphics[width=0.65\linewidth]{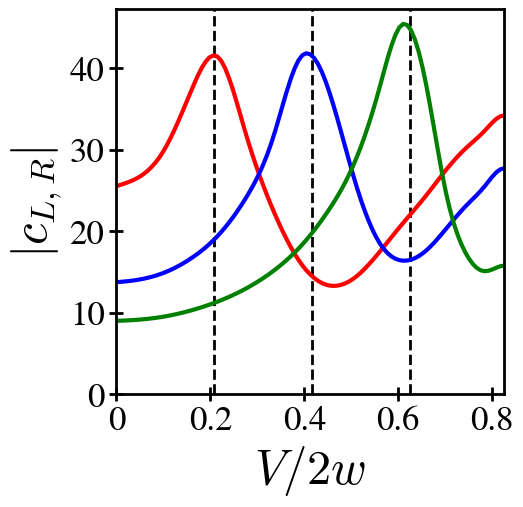}
  \caption{$|c_{L,R}|$   in the NESS, computed according to Eq. (\ref{appe.2.3}), 
  for the system described by the Hamiltonian in Eq. (\ref{appe.2.1}), with the following values of the various parameters:
$2w=1$, $\tau=0.04$,  $\epsilon_0=0.21,0.42,0.63$ (red, blue and green curve, respectively). The dashed, vertical lines
 mark the position of the values of $\epsilon_0$ we used in the calculations.  
 }
  \label{f.appe.2.2}
\end{figure}

To motivate our use of ${\cal I}$ and $C_{L,R}$ to probe the phase of the LRK connected to the leads,  we
now preliminary analyze their behavior in a simplified
 model, in which the two leads are connected to a simple system made out of two real fermionic modes, 
 with tunable overlap with each other. By extrapolation, we then generalize our conclusions to the LRK, including 
 the effects of the disorder, as well.  
Following Refs. \cite{Nilsson2008,Giuliano2018,Guerci2021}, we define the single-level lattice model Hamiltonian $H_{\rm SL}$ as
$H_{\rm SL} =   H_{\rm 0,SL} + H_{\rm \tau,SL} + H_{\gamma}$, with 

\begin{eqnarray}
H_{\rm 0,SL} &=& - w\: \sum_{i=1}^{M-1} \sum_{X=L,R} \:   \{ d_{X,j}^\dagger d_{X,j+1} + 
d_{X,j+1}^\dagger d_{X,j} \} \nonumber \\ 
 H_{\gamma} &=&  - \frac{i}{2}  \epsilon_0 \Gamma_L \Gamma_R 
\label{appe.2.1}  \\
 H_{\rm \tau,SL} &=&   - \tau \Gamma_L (d_{L,M}^\dagger - d_{L,M} ) 
+ i  \tau \Gamma_R ( d_{R,1}^\dagger +  d_{R,1} ) \nonumber
\:\: . 
\end{eqnarray} 
\noindent
In Eq. (\ref{appe.2.1}),  $d_{(L,R),j}$ are complex lattice (lead) Dirac fermion operators, $\Gamma_L , \Gamma_R$ are 
real fermionic modes, 
and $\epsilon_0$, which measures the energy splitting due to the hybridization between the Majorana modes, is our
tuning parameter.  To faithfully mimic the subgap behavior of the LRK, 
we couple the leads, described by  $H_{\rm 0,SM}$, to the reservoirs, via the jump operators 
defined in Eqs. (\ref{jump.1}) and (\ref{eq:jump_operators}). Solving the corresponding  equation for the density 
matrix, $\rho_{\rm SL} (t)$, as in Eq. (\ref{eq:Lindblad}), we eventually compute both the 
interface current ${\cal I}_{\rm SL}$ and the (modulus of the) correlations, $|c_{L,R}|$, respectively given by

\begin{eqnarray}
{\cal I}_{\rm SL} (t) &=& - \frac{ \tau}{2}  {\rm Tr} [\rho_{\rm SL} (t) \{ ( \Gamma_L - i \Gamma_R)    d_{L,M} + {\rm H.c.} \}  ]
\nonumber \\
|c_{L,R} (t)| &=& |{\rm Tr} [\rho_{\rm SL} (t ) d_{L,M}^\dagger d_{R,1} ] |
\:\:.
\label{appe.2.3}
\end{eqnarray}
\noindent 
with  H.c. standing for Hermitian conjugate. As $t \to \infty$, ${\cal I}_{\rm SL} (t)$  and $|c_{L,R} (t)|$  reach  the 
value they take  in the NESS, which is the quantity we plot in Figs.\ref{f.appe.2.1} and \ref{f.appe.2.2}, 
as a function of the voltage bias $V$,
for different, representative values of $\epsilon_0$ (see the figure caption for details). 

\noindent

\noindent

From the plots in Fig. \ref{f.appe.2.1}, we note that, at a given $\epsilon_0$,  ${\cal I}_{\rm SL}$, as a function 
of $V$, shows an   activated behavior at a  threshold  set at $V \sim \epsilon_0$. As we discuss above, 
whether the activated behavior emerges, in our system, at zero, or at a finite, bias signals 
whether the LRK is in the topological, or
in the topological massive, phase. The same information can be recovered from the plots of  $|c_{L,R}|$  as a function of 
$V$,  which we show in Fig. \ref{f.appe.2.2}. Indeed, we see that, at a given $\epsilon_0$, the curves show 
 a peak, centered around $V \sim \epsilon_0$, whose location is consistent with what we 
 get from  ${\cal I}_{\rm SL}$ as a function of $V$ in  Fig. \ref{f.appe.2.1}.
  
 Using the results of this Section as our reference starting point, we now look at the charge current and 
 the correlations in the NESS to study the phase of the LRK, in the clean limit as well as in the 
 disordered case.

\subsection{Charge current through the LRK model at the NESS}
\label{ccness}

\begin{figure}[t]
  \centering
  \includegraphics[width=1.0\linewidth]{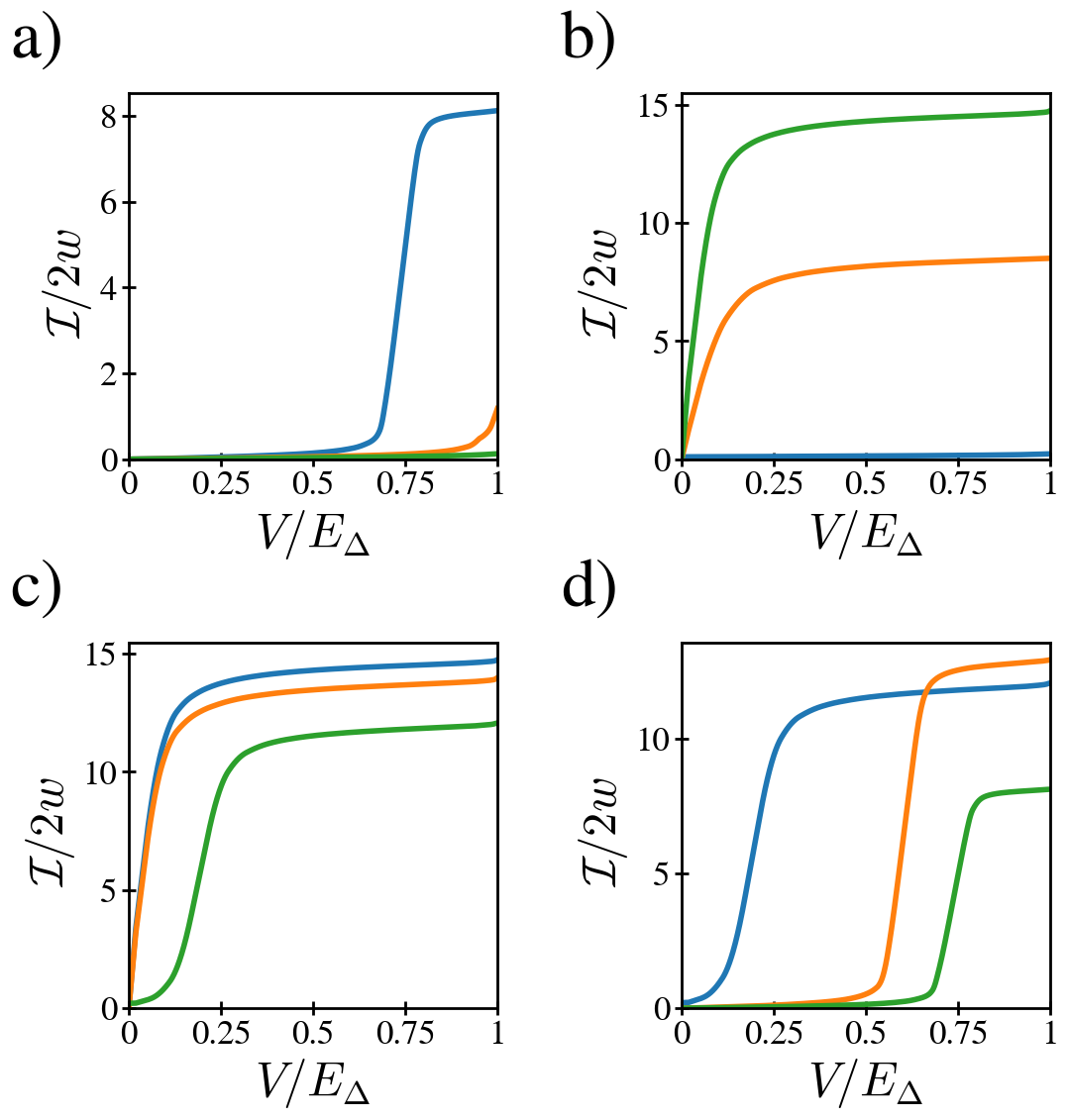}
  \caption{Charge current ${\cal I}/2w$  computed, in the absence of disorder,  for the LRK described by $H$ 
  in Eq. (\ref{eq:LRK}) with $2w=\Delta=1$ and $L=500$. The panels correspond to points over any of the 
  four segments connecting the point A,B,C,D in the phase diagram of Fig. \ref{fig:ps_clean}{(a)}. In particular:\\
  {(a)}: $\mu=-1.25$ for all the plots, $\alpha=0.5$ (blue line), $\alpha=1.25$ (orange line), $\alpha=2.5$ (green line). These
  points all lie on the AB segment of Fig. \ref{fig:ps_clean}{(a)}; \\
  {(b)}: $\alpha=2.5$,   $\mu=-1.25$ (blue line), $\mu=-0.75$ (orange line), $\mu=0.5$ (green line). These
  points all lie on the BC segment of Fig. \ref{fig:ps_clean}{(a)}; \\
  {(c)}: $\mu=0.5$,   $\alpha=2.5$ (blue line), $\alpha=1.25$ (orange line), $\alpha=0.5$ (green line). These
  points all lie on the CD segment of Fig. \ref{fig:ps_clean}{(a)}; \\  
  {(d)}: $\alpha=0.5$,   $\mu=0.5$ (blue line), $\mu=-0.75$ (orange line), $\mu=-1.25$ (green line). These
  points all lie on the DA segment of Fig. \ref{fig:ps_clean}{(a)}. 
 }
  \label{abcd_1}
\end{figure}

Referring to the formalism of Section \ref{sec:appendix_current} and of 
Appendix \ref{lindblad}, we express  the charge current ${\cal I}$ in the normal leads by means of the continuity equation for 
the particle number at site $j$, $n_j (t) = {\rm Tr} [ \rho (t) c_j^\dagger c_j]$, given by 

 \begin{eqnarray}
 && \frac{ d n_j (t)}{d t} = - 2 \sum_{l \neq j} \Im m [ A_{l,j} \theta_{l,j} (t) ] 
 \label{ccn.2} \\
 && - \sum_{\ell \neq j} 
 \Re e [({\cal G} + {\cal R} )^*_{l,j} \theta_{l,j} (t) ]  +  {\cal G}_{j,j} (1-n_j (t) ) + {\cal R}_{j,j} n_j ( t ) \nonumber 
  \:\: . 
 \end{eqnarray}
 \noindent
 Eq. (\ref{ccn.2}) can be readily regarded as a continuity equation, evidencing how there are two different 
  sources of the nonzero rate of change $\frac{ dn_j (t)}{d t}$. Specifically: 
 
\begin{itemize}
    \item The ``standard'' kinetic contribution, related to the nonzero hopping amplitude encoded in the matrix 
    $A$. The corresponding contribution to the particle current over the link between $i$ and $j$ is given by  
     \cite{Nava2023ssh}:

 \beq
J_{i,j} (t)  = - 2 \Im m \{ A_{i,j} \theta_{i,j} (t) \}    
\:\: .
\label{ccn.3}
\eneq
\noindent     

    \item The second contribution, at the  bottom line of Eq. (\ref{ccn.2}), 
  due to the reservoirs, that exchange particles with the leads at a given rate. 
\end{itemize}  
Since, in the NESS 
 $\frac{ d \Theta_{\rm NESS} (t)}{d t} = 0$, the two contributions at the right hand
side of Eq. (\ref{ccn.2}) compensate with each other, so that the  
current  ${\cal I} = J_{M,M+1}$ provides all the relevant information about subgap states
in the LRK, just as ${\cal I}_{\rm SL}$ does in the single level, simplified model. To 
evidence the similar behavior of ${\cal I}$ and ${\cal I}_{\rm SL}$, in Fig. \ref{abcd_1}, we 
show sample plots of ${\cal I}$ in the absence of disorder,  along the straight lines connecting the four points ABCD in the 
phase diagram of Fig. \ref{fig:ps_clean}{(a)}.

\noindent
 Figure \ref{abcd_1}{(a)} contains plots of ${\cal I}$ as a function of $V$, normalized to the bulk gap $E_\Delta$, 
computed for $\mu=-1.25$ and $\alpha=0.5$ (blue curve), $\alpha=1.25$ (orange curve) and $\alpha=2.5$ (green curve).
All the  three points lie over the AB segment of Fig. \ref{fig:ps_clean}{(a)}. At 
  point A, which corresponds to the massive topological phase, at $V/E_\Delta \sim 0.5$, we see an activation in   ${\cal I}$,  
  which is expected to correspond to  the position of the 
  subgap massive mode. On increasing $\alpha$ at fixed $\mu=-1.25$, the system evolves toward point
B, corresponding to the trivial phase of the LRK, with no subgap modes. The shift of the activation threshold for 
${\cal I}$ toward the bulk gap edge ($V/E_\Delta =1$) witnesses that the transition from the massive topological 
to the trivial phase comes along with a progressive shift of the subgap mode toward the bulk gap threshold. 
At the same time, moving from the blue to the orange and then to the green  curve,
the over-all curve continuously evolves toward a plot constantly equal to 0 (for $V/E_\Delta \leq 1$), which is 
perfectly consistent with the expected disappearance of ${\cal I}$ in the trivial phase in this range of values of 
$V$.  In Fig. \ref{abcd_1}{(b)}, we show plots  drawn at constant $\alpha=2.5$ and for
$\mu=-1.25$ (blue curve), $\mu=-0.75$ (orange curve) and $\mu=0.5$ (green curve), all lying 
over the BC line of Fig. \ref{fig:ps_clean}{(a)}, connecting point B, in the trivial phase, with point C, 
in the topological phase of the LRK. Tuning $\mu$ along the BC line corresponds to the ``standard'' topological
phase transition in the Kitaev model. This means that either there are no subgap modes or, once the system
has entered the topological phase, they are pinned at zero energy. At the same time, due to the resonant Andreev
reflection triggered by the localized real fermionic mode at each interface \cite{Fidkowski2011,Affleck2013,Affleck2014}, 
we expect a strong increase of the current across the phase transition. Indeed, Fig. \ref{abcd_1}{(b)} is consistent
with both expectations. On one hand, on increasing $\mu$ there is an apparent increase of ${\cal I}$ when entering 
the topological phase, at any $V/E_\Delta<1$. On the other hand, the activation threshold keeps locked at 
$V/E_\Delta \sim 0$, consistently with the emergence of the localized, zero energy  modes. The plots in 
Fig. \ref{abcd_1}{(c)} are drawn at $\mu=0.5$ and with the same values of $\alpha$ as in Fig. \ref{abcd_1}{(a)},
in decreasing order, starting from $\alpha=2.5$. Now the LRK is taken from the topological to the massive topological phase and, 
consistently, we see a shift of the activation threshold toward a finite value of $V/E_\Delta$, accompanied by a mild 
reduction in ${\cal I}$ at a given $V$. Finally, Fig. \ref{abcd_1}{(d)}  summarizes the evolution of ${\cal I}$ as a function of 
$V/E_\Delta$ when moving from point D to point A of  Fig. \ref{abcd_1}{(a)}. The shift of the activation threshold across
finite values of $V/E_\Delta$ is definitely consistent with what one expects along a line connecting two massive topological 
phases with different values of the energy of the subgap mode.  

As the above discussion evidences,   the behavior of ${\cal I}$ as a function of $V/E_\Delta$ for $V/E_\Delta<1$ 
can be readily interpreted in 
terms of an effective, single level Hamiltonian as the one in Eq. (\ref{appe.2.1}). In the presence of disorder 
we compute the current going through the following, additional steps:

\begin{enumerate}
    
    \item We randomly generate a configuration of the 
   impurity potential, $\{\epsilon_i\}$, as discussed in the previous section; 
    
    \item  We solve Eq. (\ref{def.5}) for $\Theta_{\rm NESS}$ at given $\{\epsilon_i \}$;

    \item  We repeat the two previous points   $\mathcal{N}$ times and eventually average over the realizations of the disorder.
    Accordingly, denoting with $\mathcal{I}^{(r)}$ the current computed for the $r$-th realization of the disorder,  
    we compute the   disorder-averaged current $\langle {\cal I} \rangle$ as:  
   
   \beq
   \langle {\cal I} \rangle = \frac{1}{\mathcal{N}} \: \sum_{r=1}^{\mathcal{N}} \: \mathcal{I}^{(r)} 
   \:\:.
   \label{ccn.4}
 \eneq
 \noindent

       \item Repeating the above procedure, at all the marked points A,B,C,D in
  Fig. \ref{fig:ps_clean}, we construct plots of $\langle {\cal I} \rangle$ as a function of 
  $V/E_\Delta$ for various values of the disorder strength $W$. 
\end{enumerate} 

Remarkably, previous studies of the effects of the disorder on a topological phase 
have shown that, while a small amount of disorder works to enforce the topological phase(s), 
 a strong enough value of $W$ typically washes out the topology, typically when  
it becomes of the same order of magnitude as $\mu$ 
 \cite{Pientka2013,Nava2017,Zuo2022,Li2009,Liu2022,Nava2023ssh,Cinnirella2024}.
 Yet, the specific interplay between disorder and topology depends on the characteristics of the 
reference clean phase, which, in our case,  motivates our systematic analysis  of the effects
of  increasing the disorder strength $W$ at the points 
A,B,C, and D in Fig. \ref{fig:ps_clean}.

\begin{figure}[t]
  \centering
  \includegraphics[width=1.0\linewidth]{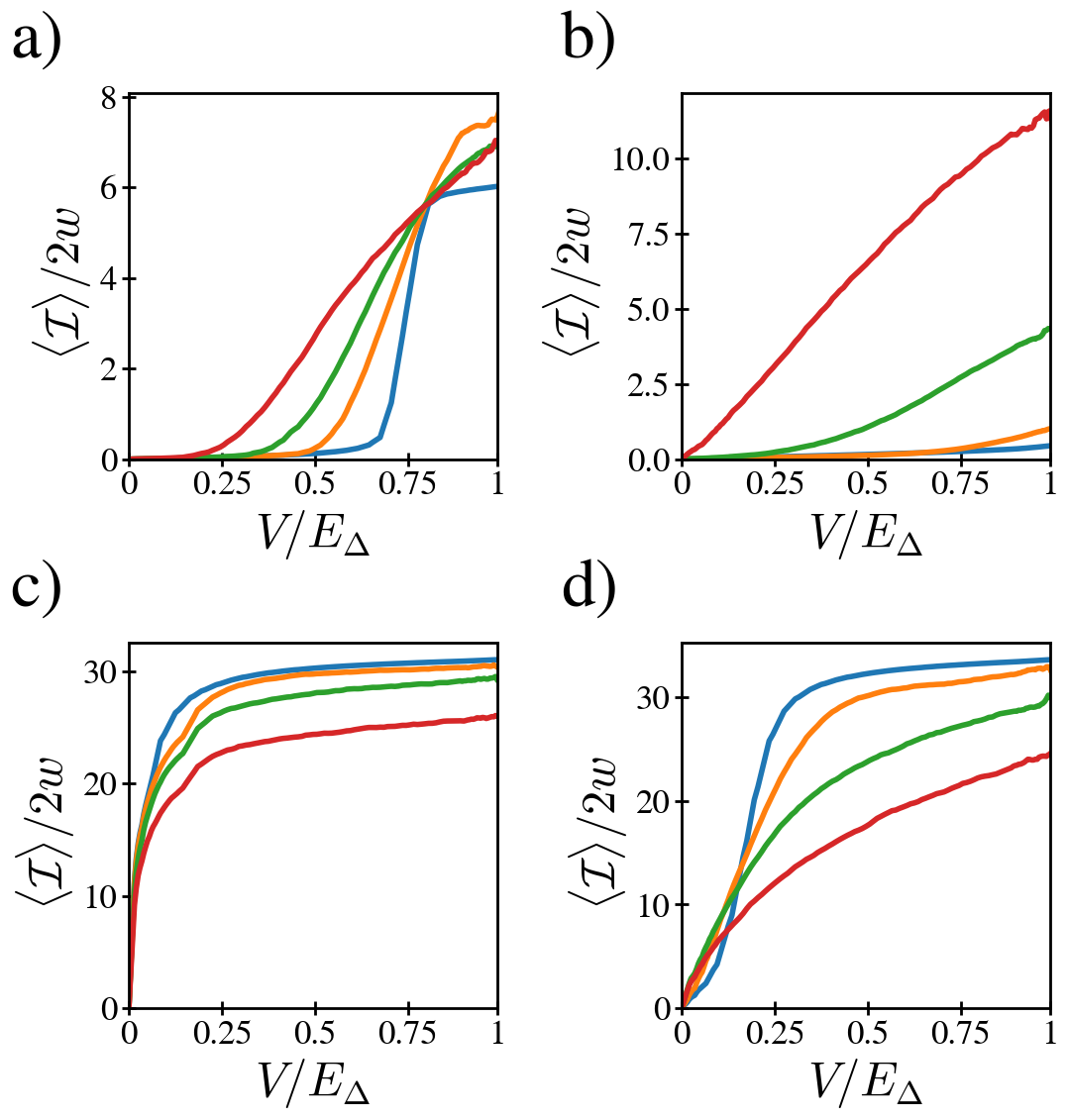}
  \caption{$\langle {\cal I}\rangle/2w$ computed  at the NESS for the LRK  
   with $2w=\Delta =1$, $L=100$, connected to metallic leads with $M=101$ 
  and variable single electron hopping strengths  
  $\gamma_R=\gamma_L=0.05$. The coupling strengths between the leads and 
  the reservoirs are set at $g=0.01$. In any case, the temperature is set to 0. The strength of the disorder is 
  varied as detailed in the following. \\
  $\langle{\cal I}\rangle/2w$ vs $V/E_\Delta$ at points (a) A, (b) B, (c) C, and (d) D of Fig. \ref{fig:ps_clean} for disorder $W=0$ (blue curve), $W=0.2$ (orange curve), , $W=0.4$ (green curve), $W=0.6$ (red curve).}
  \label{disfig_1}
\end{figure}
\noindent
Our main results are summarized by  the plots of in Fig. \ref{disfig_1}.
In particular, in Fig. \ref{disfig_1}{(a)}, we show plots of $\langle{\cal I}\rangle$ {\it vs.} $V/E_\Delta$ at point A of the phase
diagram of Fig. \ref{fig:ps_clean} at increasing values of $W$. While the blue curve, drawn at $W=0.0$, shows 
an activated behavior at a finite value of $V/E_\Delta$, corresponding to the location in energy of the subgap
mode, on increasing $W$ the activation threshold moves to lower (subgap) values energy.  Such a behavior 
is qualitatively similar to the one that we show in Fig. \ref{disfig_1}{(d)}, corresponding to point D, which is 
in the same (massive topological) phase as point A, although, due to the relatively smaller value of the 
subgap mode energy, the shift in the activation threshold is not as evident as at point A. Conversely, 
Figs. \ref{disfig_1}{(b)} and \ref{disfig_1}{(c)} are consistent with the known results for the short-range pairing, disordered Kitaev model 
 \cite{Pientka2013,Nava2017}. Specifically, the
persistence, at increasing $W$,  of the activation threshold at $V/E_\Delta =0$ in Fig. \ref{disfig_1}{(c)} 
evidences that a limited amount of disorder does not affect the topological phase and only results in a 
relatively small lowering of $\langle{\cal I}\rangle$ on increasing the disorder strength, while the absence of a well-defined 
threshold in Fig. \ref{disfig_1}{(b)} corresponds to the absence of subgap modes in the clean limit, with the
increase of $\langle{\cal I}\rangle$ on increasing $W$ possibly determined by the 
proliferation of disorder-induced subgap states. 

The shift toward zero energy, at increasing disorder, 
of the activation threshold in Figs. \ref{disfig_1}{(a)} and \ref{disfig_1}{(d)} evidences how
a limited amount of disorder enforces the localization of edge modes in the short-range Kitaev model. 
Consistently,  the finite energy subgap mode in the massive topological phase of the LRK can be thought 
of as due to the  long-range hybridization between edge modes  
\cite{Vodola2016}. Indeed, an increase in the localization of the boundary modes corresponds to a decrease in the corresponding
energy, which is perfectly consistent with the shift in the activation threshold toward lower energy on increasing $W$. 
This suggests the  possibility of realizing a disorder induced phase transition from 
the massive topological phase to   the short-range topological phase which, in the clean limit, would 
only be possible upon tuning an hardly tunable parameter, such as $\alpha$.  

\begin{figure}[t]
  \centering
  \includegraphics[width=0.6\linewidth]{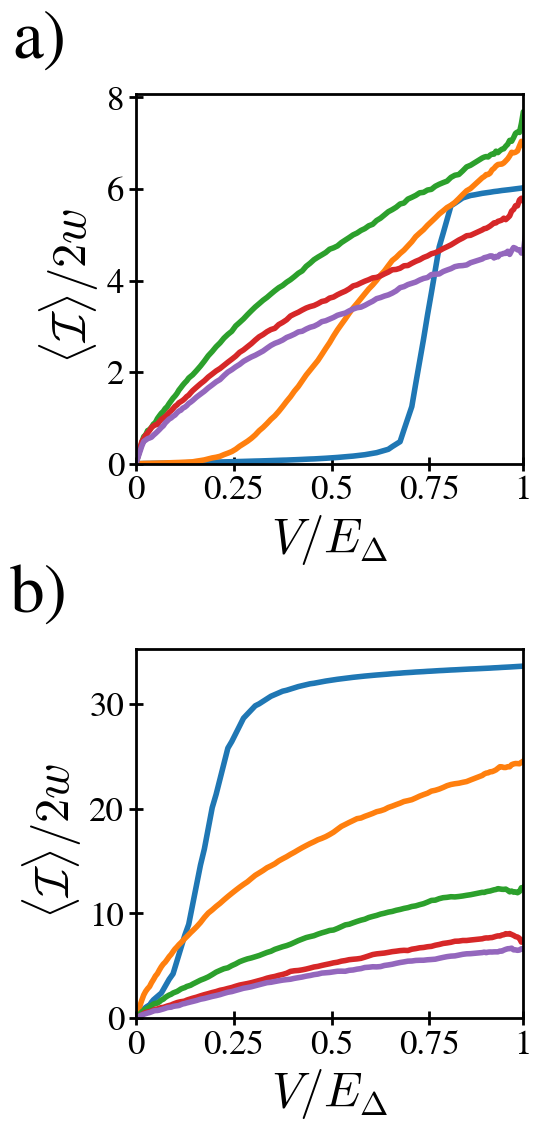}
  \caption{$\langle{\cal I}\rangle/2w$  computed  for the same system as in 
  Fig. \ref{disfig_1}{(a)} (panel {(a)}) and in Fig. \ref{disfig_1}{(d)} (panel {(b)}), 
  with  $W=0$ (blue curve),   $W=0.6$ (orange curve), $W=1.2$ (green curve), $W=1.8$ (red curve), $W=2.0$ (purple curve).
 }
  \label{largedis}
\end{figure}

To further ground our conclusions, in Fig. \ref{largedis} we plot $\langle{\cal I}\rangle$ versus $V/E_\Delta$ at points A and D, 
up to values of the disorder strength as large as $W=2.0$. In Fig. \ref{largedis}{(a)}, we note the shift of the activation threshold to 
zero energy and 
the rise of $\langle{\cal I}\rangle$ with $W$ at a given value of $V/E_\Delta$. This  is consistent with a transition toward the (short-range) topological phase,
roughly at  $W$ between 1.0 and 1.2. The subsequent decrease of the current for larger values of $W$ is, instead, consistent with the 
effects of adding disorder to the system in the topological phase. At variance, the monotonical decrease of $\langle{\cal I}\rangle$ with increasing $W$ in   
Fig. \ref{largedis}{(b)} is consistent with either a direct transition toward  the trivial phase, or with no transitions at all.

\noindent
To double check the conclusions of this section, in the following, we 
perform a similar study of the behavior of the endpoint  correlations across the LRK.

\subsection{Endpoint correlations between the leads}
\label{correlations}

\begin{figure}[t]
  \centering
  \includegraphics[width=0.7\linewidth]{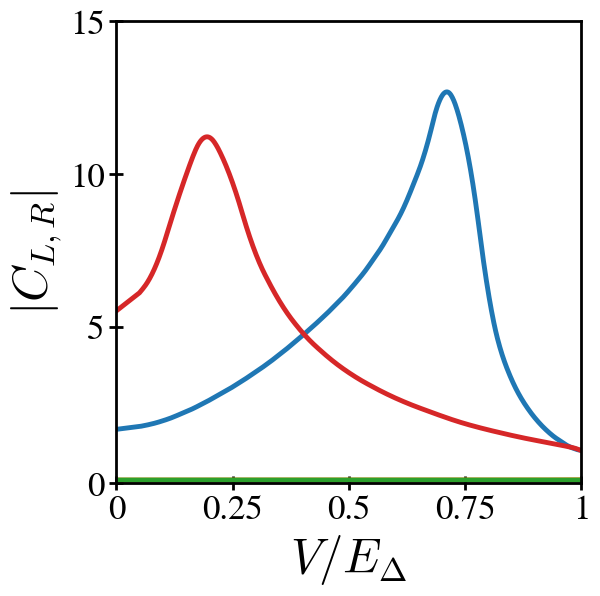}
  \caption{$|C_{L,R}|$  computed  for the same system as in 
  Fig. \ref{abcd_1}, at points A  (blue curve), B (orange curve), C (green curve), and 
  D (red curve) of Fig. \ref{fig:ps_clean}{(a)}. Note that the orange and the green curve have
  collapsed onto the $|C_{L,R}|=0$ axis. }
  \label{clean_corr}
\end{figure}

In the absence of disorder, the correlation   $C_{L,R}$ is expressed in terms of 
the covariance matrix elements as 
 
\begin{equation}
    C_{L,R}= \theta_{M,M+L+1} 
    \:\: . 
\label{eq:corr_clean}
\end{equation}
\noindent
Introducing disorder, Eq. (\ref{eq:corr_clean}) is generalized by averaging over a number 
${\cal N}$ of independent disorder realizations, that is

\begin{equation}
  C_{L,R} =\frac{1}{\mathcal{N}}\sum_{r=1}^{\mathcal{N}}\theta^{(r)}_{M,M+L+1}
    \:\: . 
\label{eq:corr_disorder}
\end{equation}
\noindent
In  Fig. \ref{clean_corr}, we show our result for  $|C_{L,R}|$, as a function 
of $V/E_\Delta$, in the absence of disorder and for $V/E_\Delta <1$, at the four points A,B,C, and D in 
Fig. \ref{fig:ps_clean}{(a)}. While, at the A and D points, $|C_{L,R}|$ shows a peak centered at an energy consistent with the position of 
the activation threshold of ${\cal I}$ in the clean limit (blue and red curve, respectively), the curves corresponding 
to points B and C, within the whole interval $0\leq V/E_\Delta < 1$, keep  smaller than the previous ones by several 
order of magnitudes: they are basically constantly equal to 0. As expected from our previous discussion of the simplified 
model, this is consistent with either the absence of subgap modes (point B), or with the lack of hybridization 
between the real fermionic modes (point C).

\begin{figure}[t]
  \centering
  \includegraphics[width=0.7\linewidth]{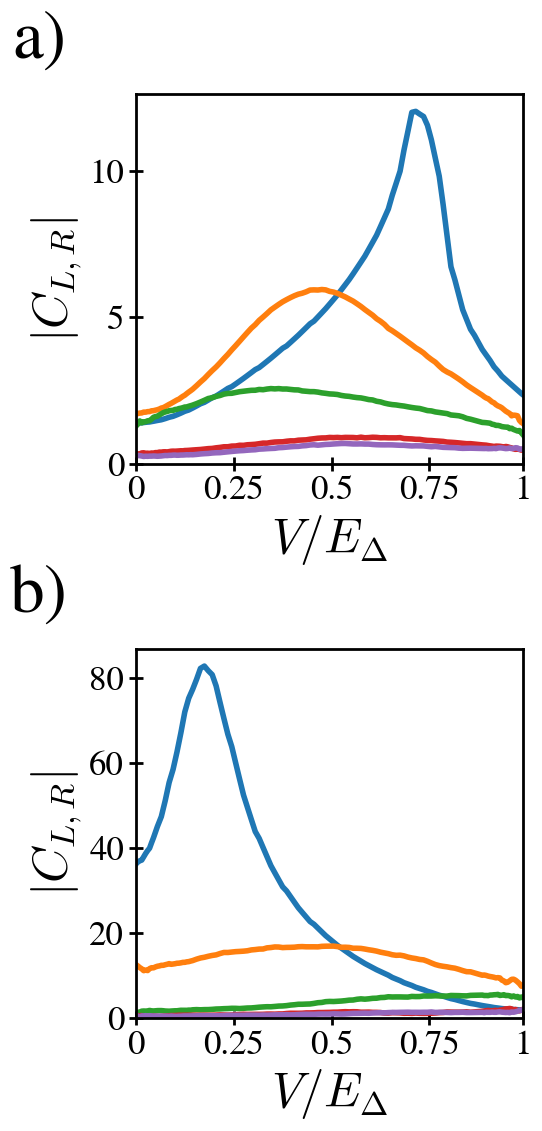}
  \caption{$|C_{L,R}|$ computed  for the same system as in 
  Fig. \ref{largedis}, at points A  (panel {(a)}) and D (panel {(b)}) of the phase diagram 
  of the LRK,  with  $W=0.0$ (blue curve),   $W=0.6$ (orange curve), $W=1.2$ (green curve),
   $W=1.8$ (red curve), $W=2.0$ (purple curve).
 }
  \label{dirti_corr}
\end{figure}

\noindent
In the disordered case, since disorder typically favors the localization, we expect 
no relevant effects at points B and C. For this reason, in the following we focus on points 
  A and D of the phase diagram. In Fig. \ref{dirti_corr}, we plot $|C_{L,R}|$ as a function of 
  $V/E_\Delta$, computed at points A and D of the phase diagram, at increasing disorder 
  strength $W$ (see figure caption for details).

\noindent
Specifically, Fig. \ref{dirti_corr}{(a)} corresponds to point A of the phase diagram, Fig. \ref{dirti_corr}{(b)} to
point D. We note that, while, in both cases, there is a strong suppression of $|C_{L,R}|$ on increasing $W$, 
in Fig. \ref{dirti_corr}{(a)} there is no inversion of the main trend, such as in Fig. \ref{largedis}{(a)}, 
at large disorder strength. In fact, as we remark above, the correlations do not allow to resolve the trivial from 
the short-range topological phase. Yet, combining the plots of Fig. \ref{dirti_corr} with the ones on 
Fig. \ref{largedis}, provides a comprehensive picture of what are the effects of the disorder in the various phases of
the LRK, suggesting, in addition,  the intriguing possibility that the disorder itself might trigger a direct transition from 
the massive to the short-range topological and, eventually, to the trivial phase. 

 \section{Concluding remarks}
\label{concl}

We have made a combined study of the phase diagram of the disordered LRK, connected to
thermal reservoirs by means of metallic leads, by first monitoring  the lowest and the 
second lowest energy eigenvalue of the LRK Hamiltonian at 
 increasing disorder strength, then, by computing the  current flowing through the interfaces between the LRK 
 and the leads at a finite bias between 
 the reservoirs  and, finally, by looking  
  at the endpoint correlations across the LRK. 
 
  Our method   has allowed us to circumvent the difficulty 
 of defining a topological invariant, such as the Chern number, in the disordered system.   
 Apart from mapping out the various phases of the disordered system, we have evidenced
 how a   limited amount of disorder (small $W$) induces reentrant massive topological phases in 
 the LRK (points B and C of the phase diagram in Fig. \ref{fig:ps_clean})  
 that are analogous to   the disorder induced reentrant topological phase of the
 Kitaev model. In particular, while a limited amount of disorder makes the system behave similarly to 
 what expected in the short-range topological phase, a strong disorder washes out the topological phase, just as in the short-range
 case. Again there is a reentrant behavior in the phase diagram, as disorder is increased, 
  but this time toward a different topological phase.  
  
 Aside from providing a first, comprehensive study of the disordered LRK, the results we obtained within our approach call for  
 an extension of our work to make, for instance, a sharp verification of whether it is possible to recover a direct  
 transition between the massive and 
 the short-range topological phase by acting on the disorder only, possibly changing the functional form of the disorder potential
 as, for instance, done in Refs. \cite{Nava2023ssh,Cinnirella2024}. Yet, this, as well as other possible developments of our results, 
 go beyond the scope of this work and will possibly be analyzed in forthcoming research projects. 
The data that support the findings of this article are openly available at \cite{cinnirella_2025_15221285}

\begin{acknowledgments}
A. N. acknowledges funding by the Deutsche Forschungsgemeinschaft (DFG, German Research Foundation),
Projektnummer 277101999 -- TRR 183 (project C01), under project No.~EG 96/13-1,
and under Germany's Excellence Strategy -- Cluster of Excellence Matter and Light for 
Quantum Computing (ML4Q) EXC 2004/1 -- 390534769. E.G.C. acknowledges funding (partially) supported by ICSC Centro Nazionale di Ricerca in High Performance Computing, Big Data and Quantum Computing, funded by European Union—NextGenerationEU.
\end{acknowledgments} 

\appendix

 \section{Lindblad Master Equation for the correlation matrix}
 \label{sec:appendix_current}
 
In this Appendix we present the main mathematical steps behind the formalism of
Sec. \ref{lindblad}.  In the following, we consider
 a generic quadratic fermion Hamiltonian ${\cal H}$, containing single-fermion pairing and hopping terms, 
 in the form 
   
\beq
 {\cal H} =\sum_{i,j} \left( A_{i,j} c^\dagger_i c_j + B_{i,j} c^\dagger_i c^\dagger_j +  B^*_{j,i} c_i c_j \right)
\;\; , 
\label{appe.1.1}
\eneq 
\noindent
with $c_i,c_j^\dagger$ being lattice fermionic operators satisfying the anticommutation algebra $\{c_i,c_j^\dagger \} = \delta_{i,j}$
(all the other anticommutators being equal to 0), 
  ${A^\dagger =A}$ and ${B^T=-B}$. Starting from Eq. (\ref{eq:Lindblad}) for the density matrix $\rho (t)$ and 
defining the Lindblad jump operators as in Eq. (\ref{eq:jump_operators}), we readily obtain the time evolution equation for 
$\theta_{i,j} = \langle c_i^\dagger c_j \rangle$, with $\langle \ldots \rangle = {\rm Tr} [\rho (t) \ldots ]$, in the form 
 
\begin{eqnarray} 
  &&  \frac{ d \theta_{i,j} (t)}{dt}  
    =  -i\left\langle \left[ c^\dagger_ic_j,{\cal H} \right] \right\rangle  \label{appe.1.2} \\
    &&    +\sum_{n=(L,R),k} \left( \left\langle (L^{\rm in}_{n,k})^\dagger c^\dagger_i c_j L^{\rm in}_{n,k}\right\rangle -\frac{1}{2}\left\langle \left\{ c^\dagger_ic_j,(L^{\rm in}_{n,k})^\dagger L^{\rm in}_{n,k}\right\}
    \right\rangle \right) \nonumber  \\
   &&   +\sum_{n=(L,R),k} \left( \left\langle  (L^{\rm out}_{n,k})^\dagger c^\dagger_i c_j L^{\rm out}_{n,k} \right\rangle- \frac{1}{2}\left\langle \left\{ c^\dagger_ic_j,(L^{\rm out}_{n,k})^\dagger
   L^{\rm  out}_{n,k}\right\} \right\rangle  \right)     \:\: .
  \nonumber 
    \end{eqnarray}
    \noindent
 Note that, in writing Eq. (\ref{appe.1.2}), we   
 refer to a specific system such as the one we introduce and discuss in Sec. \ref{int_cur}, that is, 
a central region consisting of a one-dimensional, $L$-site lattice, with  two,  
one-dimensional, $M$-site, metallic leads, attached to the endpoints of the chain. 
    To recover the explicit form of the right-hand side of Eq. (\ref{appe.1.2}), we 
    employ the following chain of identities (that are readily recovered from  the cyclic property of
the trace): 
      
\begin{eqnarray}
&& \sum_{l,m}\left\langle \left[ c^\dagger_ic_j,A_{l,m} c^\dagger_l c_m + B_{l,m} c^\dagger_l c^\dagger_m + B^*_{l,m} c_m c_l\right] \right\rangle \nonumber \\
 && = \sum_{l,m}\Big\{ A_{l,m}\left\langle \left[ c^\dagger_ic_j,c^\dagger_l c_m\right] \right\rangle \nonumber \\
 && \quad +B_{l,m} \left\langle \left[ c^\dagger_ic_j,c^\dagger_l c^\dagger_m\right]\right\rangle  \nonumber \\
 && \quad +B^*_{l,m} \left\langle \left[ c^\dagger_ic_j,c_m c_l \right] \right\rangle \Big\} \nonumber \\
 &&  = \sum_{l,m}\Big[A_{l,m}\left(\delta_{j,l}\left\langle c^\dagger_ic_m \right\rangle - \delta_{i,m} \left\langle c^\dagger_lc_j \right\rangle \right) \nonumber \\
&&  \quad + B_{l,m} \left( \delta_{j,l} \left\langle  c^\dagger_ic^\dagger_m \right\rangle -\delta_{j,m}\left\langle c^\dagger_ic^\dagger_l \right\rangle \right) \nonumber \\
&& \quad + B^*_{l,m} \left( \delta_{i,m} \left\langle c_lc_j \right\rangle -\delta_{i,l} \left\langle c_mc_j \right\rangle \right)\Big] \nonumber \\
 &&   =  -\left\{[A^T,\theta (t) ]+2i\left( \theta^A (t)  B+ B^*(\theta^A (t) )^+ \right)\right\}_{ij}
 ,
 \label{appe.1.3}
\end{eqnarray}
\noindent
with $\theta (t)$ and $\theta^A (t)$ being the matrices with elements respectively given by 
$\theta_{i,j} (t)$ and by $\theta^A_{i,j} (t)$. To pertinently rewrite the Liouvillian term, 
denoting with $d_{(L,R),j}$ the real space, single fermion operators for the leads, we
obtain

\begin{eqnarray}
    L^{\rm in}_{(L,R),k} &=& \sqrt{2\gamma f (\epsilon_k) }\sum_{j=1}^M S_{k,j}^* d^\dagger_{(L,R),j} \;\; ,\nonumber \\  
    L^{\rm out}_{(L,R),k} &=& \sqrt{2\gamma (1-f (\epsilon_k))} \sum_{j=1}^M  S_{k,j} d_{(L,R),j}
    \;\; , 
    \label{appe.1.4}
    \end{eqnarray}
    \noindent
    with  the $d$-operators defined as in Eq. (\ref{eq:conv_NSN}), and
    
\beq
S_{j,k} =     \sqrt{\frac{2}{M+1}} \: \sin (kj) 
\:\: . 
\label{appe.1.5}
\eneq
\noindent
Using Eqs. (\ref{appe.1.4}) and (\ref{appe.1.5}), we eventually obtain 

\begin{eqnarray}
   &&    \sum_{n=(L,R),k} \left( \left\langle (L^{\rm in}_{n,k})^\dagger c^\dagger_i c_j L^{\rm in}_{n,k}\right\rangle -\frac{1}{2}\left\langle \left\{ c^\dagger_ic_j,(L^{\rm in}_{n,k})^\dagger L^{\rm in}_{n,k}\right\}
    \right\rangle \right) + \nonumber  \\
   &&   \sum_{n=(L,R),k } \; 
   \biggl( \left\langle  (L^{\rm out}_{n,k})^\dagger c^\dagger_i c_j L^{\rm out}_{n,k} \right\rangle   -   \frac{1}{2}\left\langle \left\{ c^\dagger_ic_j,(L^{\rm out}_{n,k})^\dagger
   L^{\rm  out}_{n,k}\right\} \right\rangle  \biggr)  = \nonumber \\
&&      2\gamma \sum_{n=(L,R)  } \; 
 \sum_{k,i,m} \: S_{l,k}S_{m, k}f (\epsilon_k) \times \nonumber \\
 && \left( \left\langle d_{n,l} c^\dagger_i c_j d^\dagger_{n,m}
\right\rangle 
-\frac{1}{2}\left\langle \left\{ c^\dagger_i c_j ,d_{n,l} d^\dagger_{n,m} \right\}\right\rangle \right) \nonumber \\ 
&&+  2\gamma  \sum_{n=(L,R)  } \; 
 \sum_{k,i,m} \: S_{l,k}S_{m,k}(1-f (\epsilon_k) )  \times \nonumber \\
&& \left( \left\langle  d^\dagger_{n,l} c^\dagger_i c_j d_{n,m} \right\rangle -
\frac{1}{2}\left\langle \left\{ c^\dagger_i c_j ,d^\dagger_{n,l} d_{n,m} \right\}\right\rangle \right) 
\:\: .
\label{appe.1.6}
\end{eqnarray}
\noindent
As a next step, we now introduce  the matrices $G_N,R_N$, whose elements are 
given by 

\begin{eqnarray}
(G_N)_{l,m}&=& 2 \gamma \sum_k \:  S_{l,k} S_{m,k} f (\epsilon_k ) \:\: , \nonumber \\
(R_N)_{l,m}&=& 2 \gamma \sum_k \:  S_{l,k} S_{m,k} [1-f (\epsilon_k )]
\:\: . 
\label{appe.1.7}
\end{eqnarray}
\noindent
Using the matrices in Eqs.(\ref{appe.1.7}), we can eventually write the time evolution 
equation for $\theta (t)$ in a compact form, in terms of the matrices ${\cal G},{\cal R}$. These
are defined as  $(L+2M) \times (L+2M)$  square matrices  and, in a block representation,  
they are given by

\beq
{\cal G} ({\cal R}) = \left[\begin{array}{ccc} G_N (R_N) & {\bf 0} & {\bf 0} \\
{\bf 0} & {\bf 0} & {\bf 0} \\ {\bf 0} & {\bf 0} & G_N (R_N)
\end{array} \right]
\;\; . 
\label{appe.1.9}
\eneq
\noindent 
In terms of ${\cal G}$ and of ${\cal R}$, as well as of the matrices $A$ and $B$ 
introduced in Eq. (\ref{appe.1.1}), we obtain the time evolution equation for 
$\theta (t)$ in the form:

\begin{eqnarray}
    \frac{d \theta (t)}{d t} &=& 
    i[A^T,\theta (t) ]+2i\left( \theta^A (t)  B+ B^*(\theta^A (t) )^\dagger\right) \nonumber \\
    &-&\frac{1}{2}\left\{{\cal G}+{\cal R},\theta (t) \right\}+{\cal G} 
\:\: . 
\label{appe.1.10}
\end{eqnarray}
\noindent
Going through perfectly analogous steps, we recover the time evolution 
equation for $\theta^A (t)$, as well. This is given by

 \begin{eqnarray}
    \frac{d \theta^A (t)}{dt}
    &=& i\left(\theta^A (t) A+A^T\theta^A (t) \right)-2i\left(B^*\theta^T (t) +\theta B^*\right) \nonumber \\
    &-& \frac{1}{2}\left\{{\cal G}+{\cal R},\theta^A (t) \right\}+2iB^*
\:\: . 
\label{appe.1.11}
\end{eqnarray}
\noindent
Combining together Eqs.(\ref{appe.1.10}) and  (\ref{appe.1.11}), we obtain 
the time evolution equation for the matrix $\Theta (t)$ defined in the main text, taking 
into account that, in the case of the LRK connected to metallic leads,
 $A$ and $B$ take the form
 
 \begin{eqnarray}
 &&     A= \left[ \begin{array}{ccc} 
         A_{N_L} & A^{\rm hop}_{N_L,LRK} & {\bf 0} \\
        A^{\rm hop}_{N_L,LRK} & A_{LRK} & A^{\rm hop}_{LRK,N_R} \\
        {\bf 0} & A^{\rm hop}_{LRK,N_R} & A_{N_R}
    \end{array} \right] \nonumber \\
   && B= \left[ \begin{array}{ccc}
        {\bf 0} &  {\bf 0}  &  {\bf 0}  \\
         {\bf 0}  & B_{LRK} &  {\bf 0}  \\
         {\bf 0}  &  {\bf 0}  &  {\bf 0}  \end{array} \right] 
         \;\; , 
         \label{appe.1.12}
\end{eqnarray}
\noindent
with $A_{N_L},A_{N_R}$ being $M\times M$ matrices, 
$  A^{\rm hop}_{N_L,LRK},A^{\rm hop}_{LRK,N_R}$ ($ A^{\rm hop}_{LRK,N_L},A^{\rm hop}_{N_R,LRK}$) 
being $M\times L$ ($L\times M$) matrices, and $A_{LRK},B_{LRK}$ being $L\times L$ matrices.  
 
\bibliography{biblio_essh_appeal}
\end{document}